\documentclass[twocolumn,trackchanges,twocolappendix]{aastex631}
\usepackage{comment}
\usepackage{physics}
\usepackage{bm}
\usepackage{amsmath}
\usepackage{ulem}
\usepackage{soul}
\usepackage[flushleft]{threeparttable}
\setul{}{1pt}
\definecolor{tkcolor}{cmyk}{1,0,1,0.2}
\definecolor{ttcolor}{cmyk}{1,0,0,0.3}
\definecolor{mmcolor}{cmyk}{0,0.5,3.5,0}
\newcommand{\MM}[2]{\st{#1}\textcolor{mmcolor}{#2}} 

\definecolor{kkcolor}{cmyk}{0.0,0.7,0.25,0.35}

\newcommand\Alfven{Alfv\'{e}n }
\newcommand\Elsasser{Els\"{a}sser }
\def\mbf#1{\mbox{\boldmath ${#1}$}}

\submitjournal{ApJ}

\begin{document}

\title{Effect of Magnetic diffusion in the Chromosphere on the Solar Wind}

\author[0009-0006-3533-087X]{Masato Matsuoka}
\affiliation{School of Arts \& Sciences, The University of Tokyo, 3-8-1, Komaba, Meguro, Tokyo 153-8902, Japan}
\affiliation{Graduate School of Science, Nagoya University, Furo-cho, Nagoya, Aichi 464-8602, Japan}

\author[0000-0001-9734-9601]{Takeru K. Suzuki}
\email{stakeru@ea.c.u-tokyo.ac.jp}
\affiliation{School of Arts \& Sciences, The University of Tokyo, 3-8-1, Komaba, Meguro, Tokyo 153-8902, Japan}
\affiliation{Komaba Institute for Science, The University of Tokyo, 3-8-1 Komaba, Meguro, Tokyo 153-8902, Japan}

\author[0000-0003-1579-5937]{Takato Tokuno}
\affiliation{School of Arts \& Sciences, The University of Tokyo, 3-8-1, Komaba, Meguro, Tokyo 153-8902, Japan}
\affiliation{Department of Astronomy, The University of Tokyo, 7-3-1, Hongo, Bunkyo, Tokyo, 113-0033, Japan}

\author[0009-0006-8097-8966]{Kensuke Kakiuchi}
\affiliation{School of Arts \& Sciences, The University of Tokyo, 3-8-1, Komaba, Meguro, Tokyo 153-8902, Japan}


\begin{abstract}
We investigate non-ideal magnetohydrodynamical (MHD) effects in the chromosphere on the solar wind by performing MHD simulations for Alfv\'en-wave driven winds with explicitly including Ohmic and ambipolar diffusion. We find that MHD waves are significantly damped in the chromosphere by ambipolar diffusion so that the Alfv\'enic Poynting flux that reaches the corona is substantially reduced. As a result, the coronal temperature and the mass loss rate of the solar wind are considerably reduced, compared with those obtained from an ideal MHD case, which is indicative of a great importance of the non-ideal MHD effects in the solar atmosphere. However, the temperature and the mass loss rate are recovered by a small increase in the convection-originated velocity perturbation at the photosphere because of the sensitive dependence of the ambipolar diffusion and reflection of Alfv\'en waves on the physical properties of the chromosphere. We also find that density perturbations in the corona are reduced by the ambipolar diffusion of Alfv\'en waves in the chromosphere because the nonlinear generation of compressible perturbations is suppressed. 
\end{abstract}


\section{Introduction} \label{sec:intro}
In the solar atmosphere, the ratio of the magnetic pressure to the gas pressure generally increases with elevating altitude \citep{gary2001, wiegelmann2014}. Magnetic fields play a vital role in the dynamics and thermodynamics of the plasma in upper layers of the atmosphere. Key outcomes of the magnetic dominance are heating the corona and driving the solar wind. One of the plausible mechanisms that heat and accelerate the plasma is magnetohydrodynamical (MHD) waves 
\citep[][see \citet{vanDoorsselaere2020} for recent review]{alfven1947, osterbrock1961, uchida1974,ofman1995}.
Convective motions beneath the photosphere excite various modes of waves. In particular, transverse ($\approx$ Alfv\'enic) waves are considered to be reliable players in the upward transport of energy, unlike compressible waves, they can propagate a long distance owing to the incompressible nature, avoiding shock dissipation as a consequence of steepening of wave front. Recently, transverse waves have been detected in the chromosphere \citep{okamoto2011, mcintosh2011, jess2023, yuan2023}, and in the corona 
\citep{nakariakov1999, tomczyk2007, anfinogentov2015,banerjee2021}. Roles of Alfv\'{e}nic waves in the heating and/or acceleration of the coronal plasma have also been investigated from a theoretical point of view 
\citep[e.g.,][]{alazraki1971, belcher1971, ionson1978, matsumoto2018, shoda2019}.

A key is how the energy of excited Alfv\'{e}n(ic) waves is exchanged for thermal and kinetic energies in the upper atmosphere. To this end, various mechanisms for wave dissipation have been proposed.
Transverse waves are converted into compressible waves by nonlinear mode conversion 
\citep{hollweg1982, kudoh1999, suzuki2004,suzuki2005, matsumoto2012,sakaue2020} 
and parametric decay instability \citep{tenerani2017,reville2018}, and the compressible waves eventually dissipate through the formation of shock waves \citep{suzuki2002}.
Alv\'enic waves are also damped via turbulent cascade \citep{hollweg1986, matthaeus1999, verdini2007, cranmer2007, shoda2018}, phase mixing \citep{heyvaerts1983, sakurai1984, mcmurdo2023} and resonant absorption \citep{hollweg1984, okamoto2015, antolin2015}.

In the fully ionized corona and solar wind, magnetic diffusion is negligible, and then, the ideal MHD approximation can be safely adopted in these theoretical models and numerical simulations. However, the approximation is no longer valid in the photosphere and chromosphere where the temperature is too low to achieve sufficient ionization \citep{vernazza1981}. As a result, non-ideal MHD effects play a substantial role in the evolution of magnetic fields there. 
In the denser photospheric region, the dominant process is the Ohmic diffusion that stems from the resistivity due to the collision between electrons and neutrals. In the less dense chromospheric region, the primary mechanism is the ambipolar diffusion induced by the drift motion between neutrals and magnetic fields coupled with charged particles 
\citep{leake2005, khomenko2014, soler2015,martinez2023}\footnote{We note that there is a regime where Hall diffusion is significant between the regions dominated by the Ohmic and ambipolar diffusion \citep{pandey2008}, whereas we do not consider it in the current paper (see Section \ref{subsec:etas}).}. In the latter case the frictional coupling between the neutral and charged components is not perfect owing to the low-density condition, and hence, the collision between neutrals and ions is the main agent for the magnetic dissipation \citep{mestel1956,brandenburg1994, zweibel2015}. 

These non-ideal MHD effects promote the damping of MHD waves, leading to the heating of ambient gas 
\citep{DePontieu1998, khodachenko2004, popescu2021,morton2023}. 
A characteristic property is that higher-frequency waves are more significantly affected by magnetic diffusion; for example, ambipolar diffusion has a severe impact on Alfv\'{e}nic waves with frequency higher than collisional frequency between ions and neutrals \citep{soler2013}. 
%
Although these effects have been investigated in the photosphere and the chromosphere 
\citep[e.g.,][]{piddington1956,osterbrock1961,Shelyag2016},
it is poorly understood how those MHD waves that have undergone the non-ideal MHD diffusion in the low atmosphere travel to the corona and the solar wind. 
The objective of this paper is to investigate the influence of the non-ideal MHD effect on the solar wind by performing numerical simulations from the photosphere to the solar wind with a self-consistent MHD model; We study how the physical properties of the corona and the solar wind are modified in the non-ideal MHD treatment, compared with those obtained under the ideal MHD approximation.

This paper is organized as follows. In section \ref{sec:method} we explain our simulation setup. In section \ref{sec:results} and \ref{sec:discussion} we show the main results and discuss related topics. We summarize the paper in section \ref{sec:summary}.

\section{Methods} \label{sec:method}
We perform non-ideal MHD simulations in a one-dimensional (1D; hereafter) magnetic flux tube that covers from the photosphere at $r=R_\odot$ to $r=r_\mathrm{out}=40R_\odot$, where $R_\odot = 6.96 \times 10^{6}~\text{km}$ is the solar radius.
For that purpose, we extend an ideal MHD simulation model originally developed by \citet{suzuki2005, suzuki2006} for the solar wind from coronal holes.

\subsection{Flux Tube Model\label{flux tube model}}
We adopt a super-radially open flux tube \citep{kopp1976, suzuki2013} that does not change with time. Cross section $A$ is given by $A = r^2 f$, where $f$ is the filling factor\footnote{Instead of the filling factor, the expansion factor $f_\text{ex}$ is widely used for setting up the open flux tube \citep[e.g.][]{suzuki2006ApJL,cranmer2007}. $f$ and $f_\text{ex}$ are related by a simple relation $f(r) = f_0 f_\text{ex}(r)$ \citep{suzuki2013}.} modeled as
\begin{gather}
  \label{expansion factor}
  f(r) = \frac{e^{\frac{r-R_\odot -h}{\sigma}} 
  + f_0 - \qty(1-f_0)e^{-\frac{h}{\sigma}}}
  {e^{\frac{r-R_\odot -h}{\sigma}} + 1} .
\end{gather}
We set $f_0 = f(R_\odot) = 1/1265$, $\sigma = (1/2)h$ and $h=0.042~R_\odot$, where a small value of $f_0$ indicates that the solar surface is mostly occupied by closed magnetic loops. We determine $B_{r,0}=1.48~\rm kG$ (see Section \ref{subsec:boundary condition} for this specific value).

The radial component of magnetic field $B_r$ is determined by the conservation of magnetic flux:
\begin{equation}
\label{Bflux}
  B_r(r) = B_{r,0}\frac{R_\odot^{2}f_0}{r^{2}f}.
\end{equation}

\subsection{Basic Equations}
We solve non-ideal MHD equations including gravity, radiative cooling, thermal conduction, and phenomenological heating due to the turbulent cascade of Alfv\'{e}nic waves. The followings are the equations for the conservation of mass, the conservation of radial and perpendicular momentums, the conservation of energy, and the evolution of magnetic fields, respectively:

\begin{equation}
\label{continuity}
  \pdv{t} \rho + \frac{1}{r^{2}f}\frac{\partial}{\partial r}\qty(\rho v_{r}r^{2}f) = 0,
\end{equation}

\begin{equation}
\label{momentum,r}
\begin{split}
  \pdv{t}& \qty(\rho v_{r}) + \frac{1}{r^{2}f}\frac{\partial}{\partial r}\qty[\qty(\rho v_{r}^{2} + p + \frac{B_{\perp}^{2}}{8\pi})r^{2}f ] \\
  &= \frac{1}{r^{2}f}\qty(\frac{\rho v_{\perp}^{2}}{2}+p)\frac{d}{dr}r^{2}f - \rho \frac{GM_\odot}{r^{2}},
\end{split}
\end{equation}

\begin{equation}
\label{momentum,p}
\begin{split}
  \pdv{t}& \qty(\rho \bm{v}_{\perp}) + \frac{1}{r^{3}f^{3/2}}\frac{\partial}{\partial r}\qty[\qty(\rho v_{r}\bm{v}_{\perp} - \frac{B_{r}\bm{B}_{\perp}}{4\pi})r^{3}f^{3/2} ] \\
  &= \rho \bm{D}_{v_{\perp}},
\end{split}
\end{equation}

\begin{equation}
  \label{energy}
\begin{split}
    \pdv{t}&E + \frac{1}{r^{2}f}\pdv{r}\qty[\qty(\qty(E + p_{T})v_{r}
- B_{r}\frac{\bm{B}_{\perp} \vdot \bm{v}_{\perp}}{4\pi})r^{2}f] \\
  &= \frac{1}{r^{2}f}\pdv{r}\qty[\frac{\eta_{\text{tot}}}{4\pi}r\sqrt{f}\bm{B}_{\perp} \vdot \frac{\partial}{\partial r}\qty(\bm{B}_{\perp}r\sqrt{f})]\\
  &- \rho v_{r}\frac{GM}{r^{2}} + Q_{\text{rad}} + Q_{\text{cond}},
\end{split}
\end{equation}

\begin{equation}
  \label{induction}
\begin{split}
  \pdv{t}& \bm{B}_{\perp} + \frac{1}{r\sqrt{f}}\pdv{r}\qty[\qty(\bm{B}_{\perp}v_{r} - B_{r}\bm{v}_{\perp})r\sqrt{f}] \\
  &= \sqrt{4\pi \rho}\bm{D}_{b_{\perp}} +  \frac{1}{r\sqrt{f}}\frac{\partial}{\partial r}\qty[\eta _{\text{tot}}\pdv{r}\bm{B}_{\perp}r\sqrt{f}] .
\end{split}
\end{equation}
$\rho , v, p$, and $B$ are mass density, velocity, pressure, and magnetic field, respectively. 

Subscripts $r$ and $\perp$ denote radial and perpendicular components. $G$ is the gravitational constant and $M_{\odot}$ is the solar mass. A vector $\bm{a}$ in our coordinate system is expressed by these components as
\begin{equation}
\label{jouran vector}
    \bm{a} = a_{r}\bm{\hat{e}}_{r} + a_{\perp 1}\bm{\hat{e}}_{\perp 1} + a_{\perp 2}\bm{\hat{e}}_{\perp 2} ,
\end{equation}
where $\bm{\hat{e}}$ is a unit vector.

\begin{align}
E = \rho e + \frac{1}{2}\rho v^2 + \frac{B_{\perp}^2}{8\pi} \MM{}{,}
\end{align}
and 
\begin{align}
p_{\text{T}} = p + \frac{B_{\perp}^2}{8\pi}
\end{align}
are total energy density and total pressure, respectively, where $e$ is specific internal energy. The gas pressure is related with $\rho$ and $T$ through the equation of state:
\begin{equation}
    \label{EOS}
    p = \qty(\rho /\mu m_\text{u})k_\text{B}T,
\end{equation}    
where $m_\text{u}$ is the atomic mass unit, $k_\mathrm{B}$ is the Boltzman constant and $\mu$ is the mean molecular weight, respectively. $\mu$ is calculated by solving ionization and recombination balance (see Section \ref{subsec:etas}).

\begin{align}
\eta _{\text{tot}} = \eta_\text{O} + \eta _{\text{AD}}.
\end{align}
is the sum of Ohmic and ambipolar diffusivities, which are described later in Section \ref{subsec:etas}. 

$Q_{\text{cond}}$ represents conductive heating:
\begin{align}
    \label{Qcond}
    Q_{\text{cond}} = -\frac{1}{r^{2}f} \frac{\partial}{\partial r}\qty(F_c r^{2}f), 
\end{align}
where 
\begin{align}
\label{Fcond}
F_c =  \kappa_{0} T^{5/2}\frac{\partial T}{\partial r}
\end{align}
is Spitzer-H\"{a}rm-type conductive flux with $\kappa _0 = 10^{6}~\rm g~cm~s^{-3}~K^{-7/2}$ for electrons in fully ionized plasma under thermal equilibrium \citep{braginskii1965, matsumoto2014}. In weakly ionized gas with $T\lesssim 10^4$ K, the expression of equation (\ref{Fcond}) should be replaced with the conductive flux carried by neutral particles \citep{parker1953, koyama2000}. Although this correction is required below the transition region, the conduction term using the correct expression is largely dominated by other terms of equation (\ref{energy}) there. Coincidently, this is also true even if equation (\ref{Fcond}) is used owing to the steep dependence of $\kappa_0$ on temperature. Hence, equation 
(\ref{Fcond}) is used in the entire simulation domain.

$Q_{\text{rad}}$ represents radiative cooling, which is handled separately in optically thick and thin regimes \citep{suzuki2018}.
In the low-temperature, $T<T_\text{crt}=1.2\times 10^4$K, region, we adopt an empirical cooling rate based on observations of the solar chromosphere introduced by \citet{anderson1989}:
\begin{equation}
  \label{thick cooling}
  Q_{\text{rad}} = 4.5 \times 10^{9} \times \rho \times \text{min}\qty(1, \frac{\rho}{\rho_\text{crt}}),
\end{equation}
where $\rho_\text{crt}=10^{-16}$ g cm$^{-1}$ is a critical density. Equation (\ref{thick cooling}) gives $Q_\text{rad}\propto \rho$ in the high-density region to take into account the optically thick effect, which is in contrast to the normal dependence, $Q_\text{rad}\propto \rho^2$, in the low-density regime.
In the high-temperature region, $T>T_\text{crt}$, we adopt an optically thin cooling for ionized plasma:
\begin{equation}
  \label{thin cooling}
  Q_{\text{rad}} = \Lambda n n_{e} .
\end{equation}
where $n$ is the ion number density and $n_e$ is the electron number density. The cooling function, $\Lambda$, is adopted from the tabulated data by \citet{sutherland1993}.
For numerical stability, we connect these two regimes smoothly across $T=T_\text{crt}$ by interpolating $Q_\text{rad}$'s obtained from equations (\ref{thick cooling}) and (\ref{thin cooling}).

Following 
\citet[][see also \citet{shimizu2022,washinoue2022}]{shoda2018} we consider the dissipation of Alfv\'enic waves via turbulence in a phenomenological way \citep{hossain1995, cranmer2007}. $D_{v_{\perp i}}$ and $D_{b_{\perp i}}$ in equations (\ref{momentum,p}) and (\ref{induction}) denote turbulent dissipation coefficients of velocity and magnetic field amplitudes:
\begin{gather}
  \label{turbulance}
  D_{v_{\perp i}} = -\frac{c_d}{4\lambda _{\perp i}}\qty(\abs{z^{+}_{\perp i}}z^{-}_{\perp i}+\abs{z^{-}_{\perp i}}z^{+}_{\perp i}), \\
  D_{b_{\perp i}} = -\frac{c_d}{4\lambda _{\perp i}}\qty(\abs{z^{+}_{\perp i}}z^{-}_{\perp i}-\abs{z^{-}_{\perp i}}z^{+}_{\perp i}), 
\end{gather}
where
\begin{equation}
  \label{Elsasser}
  z^{\pm}_{\perp i} = v_{\perp i} \mp \frac{B_{\perp i}}{\sqrt{4\pi\rho}} \equiv  v_{\perp i} \mp b_{\perp i},
\end{equation}
is \Elsasser variables \citep{elsasser1950}. 
We set the nondimensional constant, $c_d=0.1$, following \citet{vanBandAT2017}.
$\lambda$ is the correlation length that is dependent on $r$ as
\begin{equation}
    \label{correlation length}
    \lambda (r) = \lambda_0 \frac{r}{R_\odot}\sqrt{\frac{f(r)}{f_0}}.
\end{equation}
We set $\lambda_0 = 10^3~\text{km}$. This value is based on the size of granule \citep{roudier1986, berger2001, abramenko2012}, whereas recent observation by the CoMP telescope reports a larger value of $\lambda_0=7.6 - 9.3 \times 10^3$ km \citep{sharma2023}.

\subsection{Non-ideal MHD Effects\label{subsec:etas}}
The Ohmic diffusion in the weakly ionized solar atmosphere is induced by the collision between electrons and neutrals. The corresponding diffusivity is derived \citep{spitzer1962, schmidt1966,blaes1994} as 
\begin{equation}
\label{eta}
\begin{split}
    \eta_\text{O} &= \frac{c^{2}m_{e}\nu_{en}}{4\pi e^{2}_{\text{c}}n_{e}} \\
    &\simeq 2.3 \times 10^{2} \frac{\max((1-x_\text{e}),0)}{x_\text{e}}
    \sqrt{\frac{T}{\text{K}}} 
    ~ \text{cm}^{2}~\text{s}^{-1} ,
\end{split}
\end{equation}
where $c$ is the speed of light, $e_c$ is the elementary charge, and $m_e$ is the electron mass. Subscripts $i,e$ and $n$ stand for ion, electron, and neutral species, respectively. $\nu_{en}=n_n\overline{ \sigma_{en}v_{en}}$ is the collision frequency between electrons and neutrals, where $\sigma_{en}$ and $v_{en}$ are respectively the cross section and the relative velocity between electrons and neutrals; the overline means the average over the velocity space, and $\overline{\sigma _{en}v_{en}}=8.3\times 10^{-10}~ T/$ $\rm K~cm^{3}~s^{-1}$ \citep{draine1983}. $x_\text{e}$ is the ionization degree, which is modeled below\footnote{The $\max$ function in equations (\ref{eta}) and (\ref{eAD}) is used to avoid negative $1-x_\text{e}$ because $x_\text{e}$ slightly exceeds 1 in fully ionized gas for the definition of equation (\ref{ionization ratio}). The physical origin of the $(1-x_\text{e})$ component is neutral number density \citep{khomenko2012}, which also $\rightarrow 0$ for fully ionized conditions, then, this approximated treatment can be justified.}.

The ambipolar diffusivity can be approximately calculated \citep{khomenko2012} as 
\begin{equation}
\label{eAD}
\begin{split}
    \eta_{\text{AD}} &= \frac{B^{2}(\rho_n/\rho)^2}{4\pi \chi \rho_{i} \rho_{n}} \\
    &\simeq 2.1 \times 10^{-16} \frac{\qty(B / \text{G})^{2}\max((1-x_\text{e}),0)^2}{\qty[\rho / \qty(\text{g cm}^{-3})]^2 x_\text{e}} ~ \text{cm}^{2}~\text{s}^{-1},
\end{split}
\end{equation}
where $\chi=\overline{\sigma_{in}v_{in}}/\qty(m_i + m_n)$ with $\overline{\sigma _{in}v_{in}} =1.9 \times$ $\rm10^{-9}~cm^{3}~s^{-1}$ \citep{draine1983}.

We calculate the ionization degree, following \citet[][see also \citet{hartmann1984,harper2009}]{yasuda2019}:
\begin{equation}
\label{ionization ratio}
\begin{split}
    x_\text{e} = \frac{n_e}{n_\text{H}} = \frac{n_p}{n_{\text{H}}} &+ \frac{n_{\text{He}^{+}}}{n_{\text{H}}} + 2\frac{n_{\text{He}^{2+}}}{n_{\text{H}}} \\ &+ {\sum_{j}A_{j}}\qty(\frac{R^{j}_{c1}}{R^{j}_{1c}} + 1)^{-1} .
\end{split}
\end{equation}
where $n_\text{H}, n_p, n_\text{He+}$, and $n_\text{He++}$ are number densities of the sum of neutral and ionized hydrogen, hydrogen ions, singly ionized helium ions, and doubly ionized helium ions, respectively. In addition to H and He, we take into account C, O, Na, Mg, Al, Si, S, K, Ca, Cr, and Fe.  $A_j$ denotes abundance in number density relative to H of $j$-th element, where the sum in equation (\ref{ionization ratio}) is taken for these elements. We adopt the standard solar abundances by \citet{asplund2009}. 
The number density ratios on the right-hand side are derived by assuming the equilibrium between ionization and recombination (see Section \ref{subsec:noneq_ion}). We treat the ionization of H with an approximated recipe that considers the ground and second energy levels of H atoms \citep{hartmann1984,harper2001}. The populations of He, He$^{+}$, and He$^{2+}$ are calculated from the ionization and recombination balance under the local thermodynamic equilibrium.

In the chromosphere with $T\lesssim 10^4$K where most of H and He are not ionized, the the main suppliers of lectrons are heavy elements, indicated by the last term of equation (\ref{ionization ratio}). We only consider the first ionization of these elements because, when the second ionization is taking place, the bulk of electrons are already supplied from ionized hydrogen\footnote{This is not strictly true for Ca because the second ionization potential $=11.9$ eV is slightly lower than the ionization potential ($=13.6$ eV) of H.}. 
$R_{c1}$ and $R_{1c}$ in equation (\ref{ionization ratio}) are the radiative recombination and photoionization rates, respectively; the ratio is calculated as
\begin{align}
\label{photo rad ratio}
    \frac{R^{j}_{1c}}{R^{j}_{c1}} 
    &= \frac{1}{n_e T}\qty(\frac{2\pi m_e k_{\text{B}}T}{h^2})^\frac{3}{2}
    \left[WT_{\text{eff}}e^{-h\nu_{1,0}/\qty(k_{\text{B}}T_{\text{eff}})} \right. \nonumber \\
    &\left. + W_{\text{gal}}T_{\text{gal}}e^{-h\nu_{1,0}/\qty(k_{\text{B}}T_{\text{gal}})} \right],
\end{align}
where $\nu _{1,0}$ is the frequency of the photoionization edge. In the first term of equation (\ref{photo rad ratio}) we approximated the radiation field of the sun by black body radiation with the effective temperature, $T_\text{eff}=$ 5780 K. Geometric dilution factor $W$ is defined as
\begin{equation}
\label{dilusion factor}
    W = \frac{1}{2}\qty[1 - \sqrt{1 - \qty(\frac{R_\odot}{r})^2}] .
\end{equation}
We are also considering galactic ionization in the second term of equation (\ref{photo rad ratio}), which comes from the interstellar radiation field. Following \citet{mathis1983}, we set $T_{\text{gal}} = 7500$ K and $W_{\text{gal}} = 10^{-14}$.

The mean molecular weight, which is used to calculate gas pressure via equation (\ref{EOS}), can be derived from the ionization degree, equation (\ref{ionization ratio}), as 
\begin{equation}
    \mu = \frac{1+4 A_\mathrm{He}+\sum_{j}N_j A_j}{1+A_\mathrm{He}+\sum_{j}A_j +x_e}, 
\end{equation}
where $A_\mathrm{He}$ is the abundance of He and $N_j$ is the mass number of $j$-th element.

It should be noted that we ignore the Hall term, which generates one tangential component of magnetic field from the other tangential component and the part of the ambipolar diffusion terms that requires the nonlinear coupling between the two tangential components. 
Their contributions are supposed to be smaller than the currently included terms. For a detailed explanation, see Appendix \ref{appendix}.

\subsection{Boundary Condition\label{subsec:boundary condition}}
At the inner boundary, $r = R_\odot$, we set $T_\text{eff}=~5780~\text{K}$. We adopt $\rho_0 = 2.5\times 10^{-7}~\text{g}~\text{cm}^{-3}$ from the ATLAS model atmosphere \citep{kurucz1979, castelli2003}. 
At the inner boundary, we assume that gas pressure and magnetic pressure are in equilibrium \citep{suzuki2013}:
\begin{equation}
    \label{inner p,Bp}
    \frac{8\pi p_0}{B_{r,0}^2} = 1.
\end{equation}
The gas pressure at the photosphere $p_0$ is derived from equation (\ref{EOS}) with $\rho = \rho_0$ and $T = T_\text{eff}$.
From equations (\ref{EOS}) and (\ref{inner p,Bp}), we determine $B_{r,0}$ $=1.48$ kG (Section \ref{flux tube model}).

We set vertical and horizontal velocity perturbation that drives MHD waves \citep{iijima2023} in a wide frequency band at the inner boundary,
\begin{equation}
    \label{power spectrum}
    \langle \delta v_0^{2}\rangle = \int_{\omega _{\text{min}}}^{\omega _{\text{max}}} P(\omega) d\omega ,
\end{equation}
from $1/\omega_\text{min} = 30~\text{min}$ to $1/\omega_\text{max} = 0.3~\text{min}$, where $P(\omega)$ is assumed to be propotional to $\omega^{-1}$. 

Vertical and horizontal velocities at the photosphere have been observed by Doppler technique \citep[e.g.,][]{oba2017} and feature tracking method \citep[e.g.,][]{november1988}. The obtained results exhibit a wide range from $0.37~\rm km~s^{-1}$ to $2.4~\rm km~s^{-1}$ \citep{title1989, oba2020}. In this study, as a fiducial value, we adopt $\langle \delta v_0\rangle = 1.25~\rm km~s^{-1}$ for both transverse $\delta v_\perp$ and longitudinal fluctuations $\delta v_r$, which is consistent with observational values taken by \citet{berger1998, matsumoto2010, chitta2012}. 

The outer boundary of the simulation region is set at $r=r_\mathrm{out}=40 R_{\odot}$. Above $r=r_\mathrm{out}$, the cell size, $\Delta r$, is enlarged and the domain covers up to $r\approx 80  R_{\odot}$ where we prescribe the outgoing boundary condition for mass and waves \citep{suzuki2005,suzuki2006}.

\subsection{Initial Condition}
\begin{figure}
 \centering
 \includegraphics[scale=0.7]{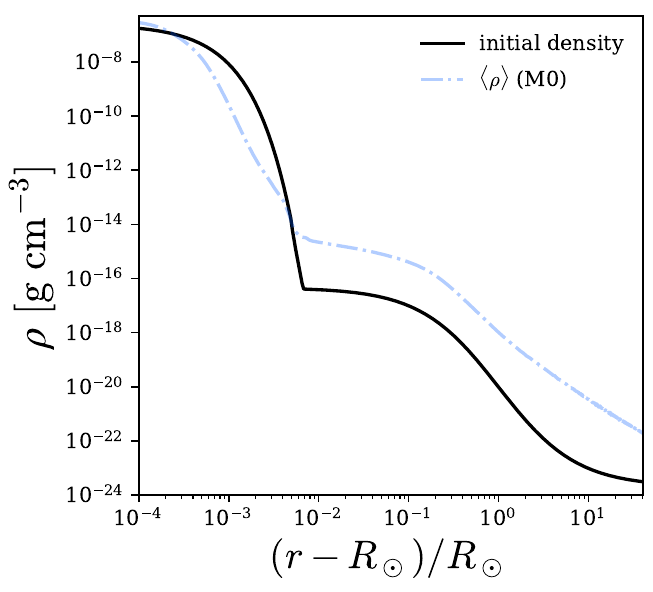}
 \caption{Initial density profile (black solid) and time-averaged radial density profile of ideal MHD case, M0 (light blue dash-dotted; see Section \ref{subsec:structures}).}
 \label{init dens}
\end{figure}

We start our simulations from the hydrostatic density structure with $T=T_\text{eff}$ in the low atmosphere where $\rho>\rho_\text{turn}=2.5\times \rm10^{-13}~g~cm^{-3}$. In the high-altitude region where $\rho < \rho_\text{turn}$, we set up higher density than the hydrostatic value to avoid unphysically high \Alfven speed, which severely limits the time step of the simulations. The initial density profile is shown in Figure \ref{init dens}.
Although the gas initially infalls from the outer overdense region, it is eventually blown outward by denser outflows from the lower region. We confirm that the final steady-state wind profile is not affected by a choice of $\rho_\text{turn}$ provided that the sufficiently small $\rho_\text{turn}$ is employed.

\subsection{Numerical Method}
The numerical scheme we adopt is the same as in \citet{shimizu2022} except that we are solving magnetic diffusion with ionizaton and recombination (Section \ref{subsec:etas}).
We update $\rho$, $\bm{v}$, $\bm{B}_\perp$, and $E$ with time by solving equations (\ref{continuity})--(\ref{induction}). The radiative cooling and thermal conduction terms in the energy equation (\ref{energy}) are updated with the time-implicit method. This is because the cooling and conduction times are short in the high-density low atmosphere and the high-temperature corona, respective, which would severely constrain the time step of the simulations when the time-explicit method was employed. 
The other terms are solved with the time-explicit method. We apply a Godunov-type shock capturing scheme \citep{vanLeer1979} that considers both gas and magnetic pressures to the compressible part of the MHD equations \citep{Sano1999,suzuki2005}. We use the method of characteristics for Alfv\'en waves \citep{Stone1992} in the incompressible part.  
We adopt second order accuracy in differentials with respect to both time and space.

\subsection{Simulation Cases\label{simulation cases}}
The main purpose of this paper is to investigate roles of magnetic diffusion in the heating and acceleration of coronal plasma. To this end, firstly we perform four cases, M0 -- M3, presented in Table \ref{energy balances} for the fixed $\langle \delta v_0\rangle$ in Section \ref{subsec:xeandRm}, \ref{subsec:structures}, \ref{subsec:energetics}, and \ref{subsec:waves}; 
in each case resistivity and ambipolar diffusion are respectively switched on and off. In addition, we also examine the dependence on the input velocity perturbation, $\langle \delta v_0\rangle$, in Section \ref{subsec:delta v0}.

We conduct the simulations until $t = 6t_\text{sim}$, where $t_\text{sim} = R_\odot / c_{s,0}$ is time in simulation units with the sound speed, $c_{s,0}$, at the photosphere.
We verify that the simulation time is sufficiently long because after $t\gtrsim 3 t_\text{simu}$ quasi time-steady profiles are achieved in the atmosphere and the wind region. We note that the simulation time, $6 t_\text{simu}$, corresponds to 10 times the Alfv\'en crossing time, $40R_{\odot}/\langle v_\text{A} \rangle$, over the simulation domain, where $\langle v_\text{A} \rangle \approx 477$ km s$^{-1}$ is the average Alfv\'en velocity from the corona to the solar wind. 

In Sections \ref{sec:results} and \ref{sec:discussion} we compare various physical quantities averaged over time. We express $\langle A \rangle$ for the average of a quantity, $A$, from $t=3t_\mathrm{simu}$ to $6t_\mathrm{simu}$.

\subsection{Energetics Formulation \label{subsec:energetics formulation}}
Under the quasi-steady state, energy balance (equation \ref{energy}) is reduced to
\begin{equation}
\label{flux conservation}
\begin{split}
    \frac{d}{dr}&\qty(L_\text{A} + L_\text{K} + L_\text{E} - L_\text{G} - L_\text{C} - L_\text{D} + L_\text{R}) \\
    &\equiv \frac{d}{dr}L_{\text{tot}} \approx 0 ,
\end{split}
\end{equation}
where total energy luminocity $L_\text{tot}$ is composed of Alfv\'en luminocity $L_\text{A}$, kinetic luminocity $L_\text{K}$, enthalpy luminocity $L_\text{E}$, gravitational luminocity $L_\text{G}$, conductive luminocity $L_\text{C}$, diffusive luminocity $L_\text{D}$, and radiative loss $L_\text{R}$ \citep{suzuki2013, shimizu2022}. They are expressed as follows:
\begin{equation}
\label{LA}
    L_{\text{A}} = \qty[v_{r}\qty(\rho \frac{v^{2}_{\perp}}{2} + \frac{B^{2}_{\perp}}{4\pi}) - B_{r}\frac{v_{\perp}B_{\perp}}{4\pi}] 4\pi r^{2}f, 
\end{equation}
\begin{equation}
\label{LK}
    L_{\text{K}} = \frac{1}{2}\rho v_{r}^{3}4\pi r^{2}f, 
\end{equation}
\begin{equation}
\label{LE}
    L_{\text{E}} = \frac{\gamma}{\gamma -1}pv_{r}4\pi r^{2}f,
\end{equation}
\begin{equation}
\label{LG}
    L_{\text{G}} = \rho v_{r}\frac{GM_\odot}{r}4\pi r^{2}f = \dot{M}\frac{GM_\odot}{r}, 
\end{equation}
\begin{equation}
\label{LC} 
    L_{\text{C}} = \kappa _{0}T^{5/2}\pdv{T}{r}4\pi r^{2}f, 
\end{equation}
\begin{equation}
\label{LD}
    L_{\text{D}} = \qty[\frac{\eta_{\text{tot}}}{4\pi r\sqrt{f}}\bm{B}_{\perp} \vdot 
    \pdv{r} \qty(\bm{B}_{\perp}r\sqrt{f})]4\pi r^{2}f, 
\end{equation}
and
\begin{equation}
\label{LR}
    L_\text{R} = -\int^{r_\text{out}}_{r}4\pi r^2 f Q_{\text{rad}},
\end{equation}
where the factor, $r^2 f$, is included to compensate the adiabatic expansion effect and
\begin{equation}
    \label{mass loss rate}
    \dot{M} = 4\pi r^2 f\rho v_r .
\end{equation}
is the mass loss rate by winds. Radiation loss, $L_{\text{R}}$, in equation (\ref{LR}) at $r$ is evaluated by the integration from $r$ to $r_\mathrm{out}$.

The sum of $L_\mathrm{A}$ and $L_\mathrm{D}$ is originally from the radial component of Poynting flux: 
\begin{align}
    \frac{L_\mathrm{A} + L_\mathrm{D}}{4\pi r^2 f} &= \frac{1}{4\pi}(\mbf{E\times B})_r \nonumber \\
    &= \frac{1}{4\pi c} \left[ (-\mbf{v\times B} + \eta_\mathrm{tot}\mbf{\nabla\times B})\mbf{\times B} \right]_r .
    \label{eq:Poynting}
\end{align}
The first term, which indicates the Poynting flux carried by Alfv\'{e}nic waves, can be separated into the outgoing component, 
\begin{align}
L_\mathrm{A,+} = \rho \qty(z_\perp^+)^2 \qty(v_r + v_\text{A}) \pi r^2 f, 
\label{eq:LA+}
\end{align}
and the incoming component, 
\begin{align}
L_\mathrm{A,-} = \rho \qty(z_\perp^-)^2 \qty(v_r - v_\text{A}) \pi r^2 f.
\label{eq:LA-}
\end{align}
We note that $L_\mathrm{A}=L_\mathrm{A,+} + L_\mathrm{A,-}$ is satisfied; $L_\mathrm{A}$ indicates the net outgoing luminocity.
The second term of equation (\ref{eq:Poynting}) is the "diffusive" (or "dissipative") Poynting flux arising from magnetic diffusion.

\section{Results} \label{sec:results}

\subsection{Magnetic Diffusivity}\label{subsec:xeandRm}

\begin{figure}
 \centering
 \includegraphics[scale=0.65]{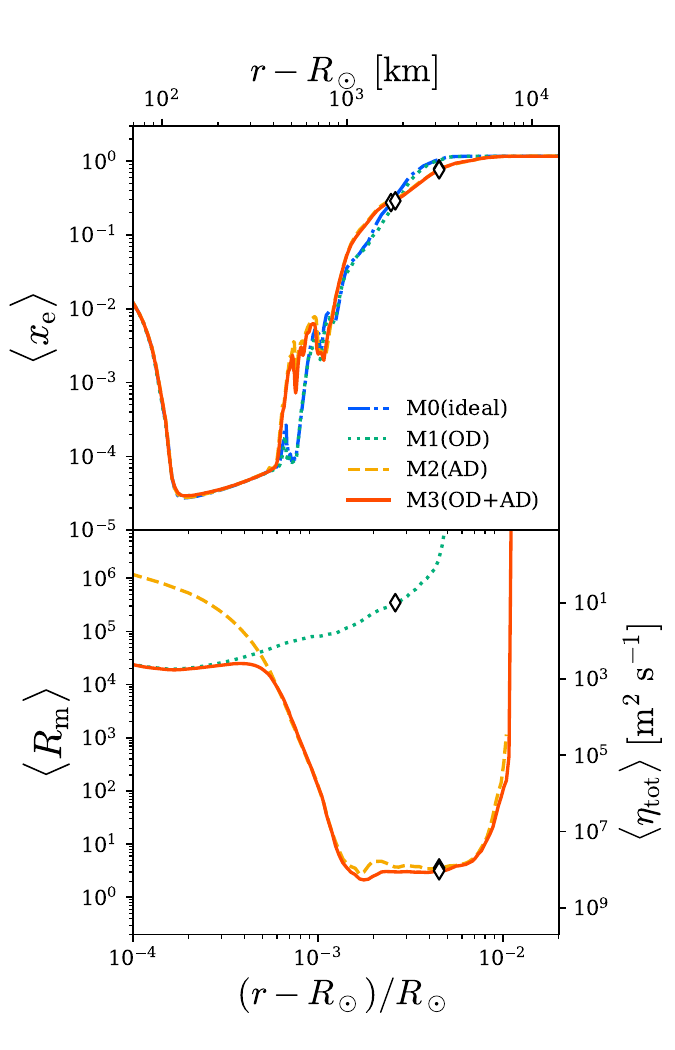}
 \caption{Time-averaged ionization fraction $x_\text{e}$ and magnetic Reynolds number $R_\text{m}$ of cases with M0 (blue dash-dotted), M1 (green dotted), M2 (orange dashed), and M3 (red solid). "OD" and "AD" stand for Ohmic and ambipolar diffusion, respectively. These panels share horizontal axis; top and bottom axes are in units of km and $R_{\odot}$, respectively. Diamonds represent the location where $T=2\times 10^{4}~\text{K}$.}
 \label{xeandRm}
\end{figure}

Figure \ref{xeandRm} shows the time-averaged radial profile of ionization fraction $x_\text{e}$ (top) and magnetic Reynolds number $R_\text{m}$ (bottom), defined below, in the low atmospheric region.
Diamond markers correspond to the location where the time-averaged temperature reaches 20000 K, which corresponds to the top of the chromosphere and the bottom of the transition region. Below this point the plasma is partially ionized and non-ideal MHD effects are non-negligible. In $100$ km  $\lesssim r \lesssim 400$ km, the ionization degree is kept small, $x_\text{e}<10^{-5}$, because the temperature there is $4000 - 5000$ K (see Section \ref{subsec:structures} and top panel of Figure \ref{1Msun_structures}) so that hydrogen is not ionized; in this region elements with low first ionization potential, such as Na and K, are the only ionization sources. Above $r-R_{\odot}\gtrsim 400$ km ($\approx 6\times 10^{-4}R_{\odot}$), $x_\text{e}$ increases with height as the temperature gradually increases in the chromosphere. Fully ionized condition is satisfied in and above the transition region.

We defined magnetic Reynolds number,
\begin{equation}
    \label{Rm}
    R_\text{m} = \frac{VL}{\eta_\text{O}+\eta_\text{AD}},
\end{equation}
where $V=10$ km s$^{-1}$ and $L=100$ km are employed for typical velocity and spatial scales, respectively, following \citet{khomenko2012}; these values roughly correspond to the sound velocity and the pressure scale height in the chromosphere. We note that the magnetic diffusivity is exactly inversely proportional to $R_\text{m}$ for constant $V$ and $L$.

In the bottom panel of Figure \ref{xeandRm}, we focus on $R_\text{m}$ of M3, which includes both Ohmic and ambipolar diffusion, as the comparison of the three cases, M1-M3, indicates that the total diffusivities, $\eta_\text{O} + \eta_\text{AD}$, in M3 is almost equal to the sum of $\eta_\text{O}$ in M1 and $\eta_\text{AD}$ in M2, which respectively include either Ohmic or ambipolar diffusion. In the photospheric region, $r-R_\odot\lesssim 500$ km, where the density is high, the Ohmic diffusion dominates the ambipolar diffusion. 
However, $R_\text{m}$ exceeds $10^4$, which means that the magnetic diffusion is not substantial there. In $r-R_\odot \gtrsim 500$ km, the ambipolar diffusion dominates and $R_\text{m}$ decreases with height as $\eta_\text{AD}(\propto x_\text{e}^{-1}\rho^{-2})$ increases, because the increase in $x_\text{e}$ (top panel of Figure \ref{xeandRm}) is overwhelmed by the rapid decrease in the density (middle panel of Figure \ref{1Msun_structures}). $R_\text{m}$ reaches the minimum value $R_\text{m}= 1-10$ at $r \approx 10^3~\rm km$ and stays $R_\text{m}<10$ in the middle and upper chromosphere, $r-R_\odot\lesssim 5000$ km ($\approx 1.007R_{\odot}$). In this region, the ambipolar diffusion significantly affects the propagation and dissipation of MHD waves, which will be discussed in the rest of the paper. At the transition region, $R_\text{m}$ jumps up as the plasma becomes fully ionized so that the ideal-MHD condition is fulfilled in and above the corona. These properties of  $R_\text{m}$ from the photosphere to the corona is consistent with what are obtained in previous works \citep{khomenko2012, martinez2012}.


\subsection{Wind Structures\label{subsec:structures}}

\begin{figure}
 \centering
 \includegraphics[scale=0.76]{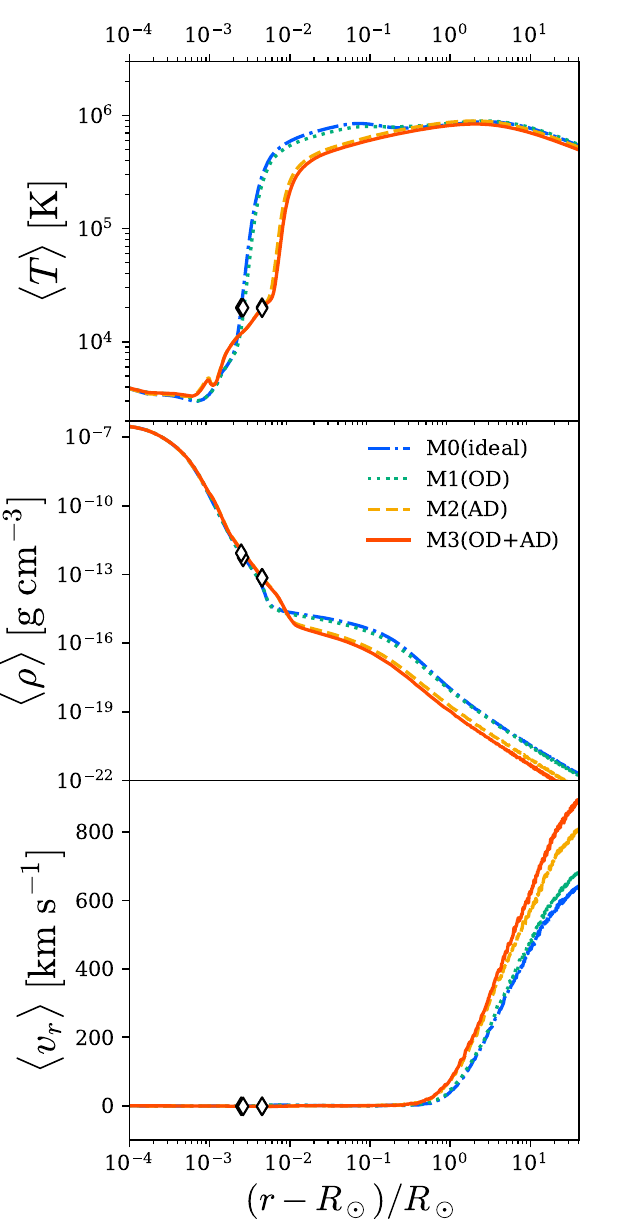}
 \caption{Time-averaged wind structures of the four cases, M0-M3. The line types and colors are the same as in Figure \ref{xeandRm}. From top to bottom, temperature, density, and radial velocities are presented.  Diamonds represent the location where $T=2\times 10^{4}~\text{K}$.}
 \label{1Msun_structures}
\end{figure}

Figure \ref{1Msun_structures} shows the time-averaged radial profiles of temperature (top), density (middle), and radial velocity (bottom). Substantial differences from the ideal MHD case (M0) are obtained when ambipolar diffusion is included (M2 and M3). The effect of the Ohmic diffusion (M1) is almost negligible except for a small difference in the radial velocity in the outer region , $r>10R_\odot$.

The cases with ambipolar diffusion give slightly higher temperature in the chromosphere (top panel of Figure \ref{1Msun_structures})
because of ambipolar diffusive heating; the transverse waves excited from the photosphere are partially damped by ambipolar diffusion in the chromosphere, which transfers the wave energy to heat \citep{khomenko2012}. The chromosphere is primarily heated by the dissipation of compressible waves \citep{arber2016}; the longitudinal waves excited in the photosphere are steepen into shock waves in the chromosphere and eventually dissipate to heat up the gas. The dissipation of transverse waves by the magnetic diffusion works as additional heating mechanism to the shock dissipation.

The effective dissipation in the chromosphere reduces the Poynting flux carried by the Alfv\'enic waves reaching the corona. As a result, the coronal heating is suppressed in the cases with ambipolar diffusion, giving lower temperature in the low coronal region of M2 and M3. The lower temperature there reduces the downward thermal conductive flux. Therefore, the evaporation of denser chromospheric gas to the corona is suppressed, and consequently, the transition region that divides the chromosphere and the corona is located at a higher altitude (diamonds in Figure \ref{1Msun_structures}), leading to the lower density at the coronal base. 
Accordingly, the density in the corona and the solar wind is also lower in these cases (M2 and M3) than that obtained in the cases without ambipolar diffusion (M0 and M1) (middle panel of Figure \ref{1Msun_structures}). The mass loss rate of the former cases is also smaller as shown in Table \ref{energy balances} (see also Section \ref{subsec:energetics}).

In the coronal region above $r-R_\odot\gtrsim 0.5 R_\odot$, the temperatures of the four cases converge. While in the cases with ambipolar diffusion, the Poynting flux reaching the corona is smaller (Section \ref{subsec:energetics}), at the same time the density is also lower. Hence, sufficient heating rate per mass is achieved to reach $T\gtrsim 10^6$ K even in these cases.

The bottom panel of Figure \ref{1Msun_structures} indicates that the wind speed of all the cases reaches several hundred km s$^{-1}$ near the outer boundary, which is an order of the escape velocity $\approx 620$ km s$^{-1}$ from the Sun. A closer inspection shows the anti-correlation between the final wind velocity and the density; it is easier to accelerate less dense wind to higher velocity. 

We would like to emphasize that, even though the non-ideal MHD effects are only important below the transition region, they make a considerable impact on the corona and the wind as shown in Figure \ref{1Msun_structures}. We examine detailed properties of the propagation and dissipation of waves in the presence of magnetic diffusion below.

\subsection{Energetics\label{subsec:energetics}}

\begin{figure*}
 \centering
 \includegraphics[scale=0.95]{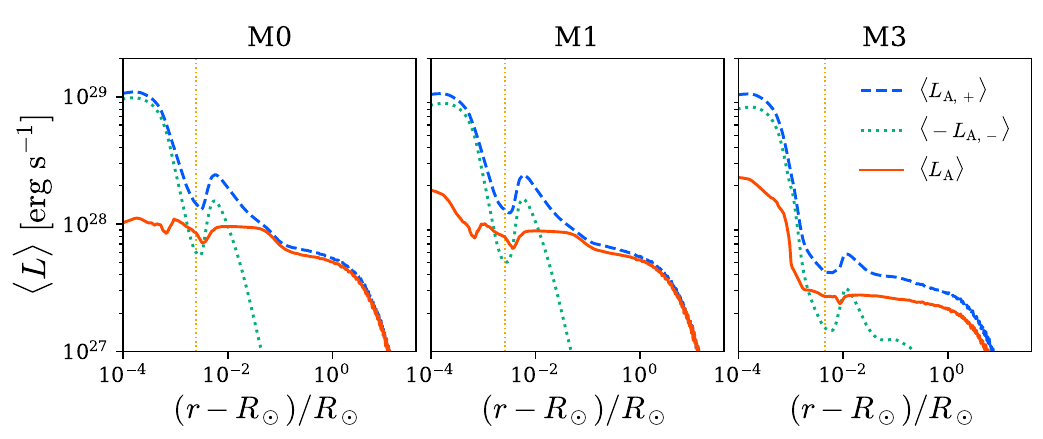}
 \caption{Time averaged radial profile of $L_\mathrm{A,+}$ (blue dashed), $L_\mathrm{A,-}$ (green dotted) and $L_\mathrm{A}$ red solid, for cases of M0 (left), M1 (middle), and M3 (right). Vertical dotted orange line represents where $T=2\times 10^4$ K.}
 \label{LA pms}
\end{figure*}

Figure \ref{LA pms} compares Alfv\'enic luminocities of M0 (left), M1 (middle), and M3 (right). In each panel, $L_\mathrm{A,+}$, $L_\mathrm{A,-}$ and $L_\mathrm{A}$ (equations \ref{eq:LA+}, \ref{eq:LA-} and \ref{LA}) are plotted. 
The outgoing Alfv\'enic luminocity at the solar surface,  
\begin{equation}    
    \label{input energy}
    L_{\text{A,}0} = -\qty(B_r\frac{\langle v_\perp B_\perp\rangle}{4\pi})_{r=R_\odot} 4\pi R_\odot^2 f_0, 
\end{equation}
in Table \ref{energy balances} is evaluated from the numerical data at 6 km (= average of four grid points from the inner boundary) above the inner boundary to avoid the effect of the boundary condition. M0--M3 yield $L_\mathrm{A,+,0}\approx (9.0-9.1)\times 10^{28}$ erg s$^{-1}$ with the difference among the four cases being less than 1\%. We note that these values are smaller than the value estimated from $\rho \langle\delta v_0^2\rangle v_\mathrm{A,0}4\pi R_{\odot}^2 f_0\approx 1.5\times 10^{29}$ erg s$^{-1}$. This is because the only velocity perturbation is input from the photosphere without magnetic fluctuation; both outgoing and incoming Poynting fluxes are injected, giving the smaller $L_\mathrm{A,+,0}$ than the simple estimate.

In the presented three cases, the incoming component, $L_\mathrm{A,-}$, (green dotted in Figure \ref{LA pms}) follows the outgoing component, $L_\mathrm{A,+}$, (blue dashed) with a slightly smaller level. This indicates that a large fraction of the injected outgoing component is reflected back downward \citep{moore1991,suzuki2006}. The comparison between $L_\mathrm{A,+,0}=9.1\times 10^{28}$ erg s$^{-1}$ of M0 in Table \ref{energy balances} and the net outgoing luminocity, $L_\mathrm{A}= 1.0 \times 10^{28}$ erg s$^{-1}$, of the same case in the photosphere (red solid line in the left panel of Figure \ref{LA pms})  illustrates that about 89\% of the input Alfv\'enic Poynting flux is reflected back to the photosphere. The reflection fractions of the dissipative cases, M1 and M3, are a little smaller but are still large, $\approx 79\%$ and, $\approx 74\%$, respectively.

The radial distribution of the Alfv\'enic luminocities in M1 is different from that of M0 only in the photosphere and the low chromosphere, $r-R_{\odot}<10^{-3}R_{\odot}$, where the Ohmic resistivity is non-negligible. The incoming mode is slightly more suppressed than the outgoing one there to give the larger net outgoing luminocity, $L_\mathrm{A}$, (red solid line in the middle panel of Figure \ref{LA pms}) near the inner boundary. On the other hand, $L_\mathrm{A}$ and $L_\mathrm{A,\pm}$ of M3 show a rapid drop at $r-R_\mathrm{\odot} \approx 10^{-3} R_\odot$ in the chromospheric region owing to the efficient ambipolar diffusion. As a result, the Alfv\'enic luminocity that reaches the transition region, $L_\mathrm{A,tc}$, of M3 is about $\approx 1/3$ of that of M0 (Table \ref{energy balances}), where "tc" stands for the top of the chromosphere at $T=2\times 10^4$ K.

\begin{figure}
 \centering
 \includegraphics[scale=0.65]{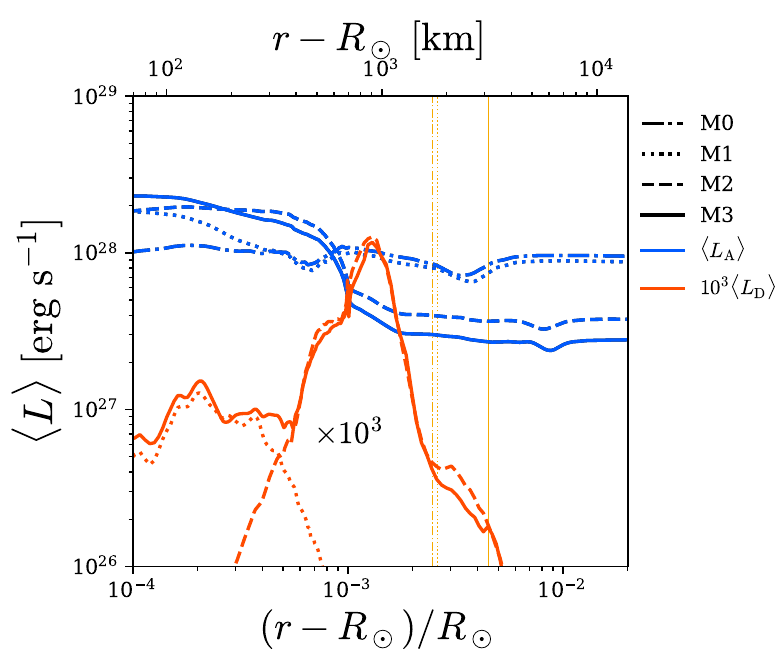}
 \caption{Time-averaged $L_\text{A}$ (blue) and $L_\text{D}$ (red) of M0--M3 in the chromosphere, where $L_\text{D}$ is multiplied by $10^3$ to fit within the displayed range. The line types are the same as in Figure \ref{xeandRm}. Orange vertical lines represent the location of the top of the chromosphere at $T=2\times 10^{4}~\text{K}$.}
 \label{LAandLD}
\end{figure}

In order to examine the dissipation of Alfv\'enic waves in the chromosphere, we show the diffusive Poynting luminocity, $L_\mathrm{D}$, (red) in addition to $L_\mathrm{A}$ (blue), below the low corona in Figure \ref{LAandLD}. We note that $L_\mathrm{D}$ is multiplied by a factor of 1000 to fit within the vertical range of Figure \ref{LAandLD}. The peaks of $L_\mathrm{D}$ at $r-R_{\odot}=2\times 10^{-4}R_\odot$ in M1 and M3 and at  $r-R_{\odot}=1.5\times 10^{-3}R_\odot$ in M2 and M3 are due to Ohmic and ambipolar diffusion, respectively. Around these peaks, $L_\mathrm{A}$ of the corresponding cases rapidly decreases, as $L_\mathrm{A}$ is converted to $L_\mathrm{D}$ there; the magnetic diffusion plays an essential role in the dissipation of the Alfv\'enic waves. However, we should note that the value of $L_\text{D}$ is much smaller than that of $L_\text{A}$. This is because the excited $L_\mathrm{D}$, which consists of the diffusive part of electric field (equation \ref{eq:Poynting}), is almost instantly converted to heat and eventually lost by radiative cooling.

\begin{table*}
  \centering
  \caption{Input parameters and time-averaged output values.}  
    \label{energy balances}  
  \begin{tabular}{cccccccccc} \hline \hline
   model& non-ideal MHD effects & $\langle \delta v_0 \rangle$ & $L_{\text{A},+,0}$ &$L_{\text{A,tc}}$ & $L_{\text{A,out}}$ & $L_{\text{K,out}}$ & $L_{\text{R,tc}}$ & $L_{\text{G,tc}}$ &  $\dot{M}$ \\ 
   & & ($\rm km~s^{-1}$) &  \multicolumn{6}{c}{($10^{27}~\rm erg~s^{-1}$)} & ($\text{M}_{\odot}~\text{yr}^{-1}$)\\
   \hline 
   M0 & $\eta_\text{O} = \eta_{\text{AD}} = 0$ & 1.25 & 91.1 & 8.49 & 0.19 & 2.77 & 3.41 & 2.45 & $2.04\times 10^{-14}$\\
   M1 & $\eta_\text{O} \neq 0$, $\eta_{\text{AD}}=0$ & 1.25 & 90.7 & 7.79 & 0.21 & 2.85 & 2.51 & 2.21 & $1.83\times 10^{-14}$ \\
   M2 & $\eta_\text{O} = 0$, $\eta_{\text{AD}}\neq0$ & 1.25 & 90.6 & 3.66 & 0.13 & 1.27 & 0.79 & 0.70 & $5.82\times 10^{-15}$\\
   M3 & $\eta_\text{O} \neq 0$, $\eta_{\text{AD}}\neq0$ & 1.25 & 90.4 & 2.70 & 0.11 & 0.94 & 0.60 & 0.42 & $3.52\times 10^{-15}$ \\ 
   M3-149 & $\eta_\text{O} \neq 0$, $\eta_{\text{AD}}\neq0$ & 1.49 & 128 & 8.39 & 0.20 & 2.95 & 2.05 & 2.41 & $2.01\times 10^{-14}$ \\
   M3-170 & $\eta_\text{O} \neq 0$, $\eta_{\text{AD}}\neq0$ & 1.70 & 166 & 13.0 & 0.19 & 3.65 & 5.40 & 4.30 & $3.57\times 10^{-14}$ \\
   \hline
  \end{tabular} 
    \begin{tablenotes}        
        \item {\bf Note.} The luminocity of each component is explained in Section \ref{subsec:energetics formulation}. The subscript "tc" or "out" indicates that the corresponding $L$ is evaluated at $r=r_\mathrm{tc}$ or  $r=r_\mathrm{out}(=40R_\odot)$.Mass loss rate $\dot{M}$ is evaluated at $r=r_\text{out}$.
    \end{tablenotes} 
\end{table*}

The Alfv\'enic Poynting luminocity that survives at the transition region basically determines the available energy to heat the corona and drive the wind. The key is that larger $L_\mathrm{A,tc}$ results in larger density at the coronal base (middle panel of Figure \ref{1Msun_structures}) because larger heating by the dissipation of Alfv\'enic waves in the corona induces more efficient chromospheric evaporation (Section \ref{subsec:structures}). Consequently, the kinetic energy luminocity, $L_\mathrm{K,out}$ ($\propto \rho v_r^3$; equation \ref{LK}), and the mass loss rate, $\dot{M}$ ($\propto \rho v_r$; equation \ref{mass loss rate}), are mostly correlated with $L_\mathrm{A,tc}$ as shown in Table \ref{energy balances} whereas the detailed dependences of $L_\mathrm{K,out}$ and $\dot{M}$ on $L_\mathrm{A,tc}$ are a little different because the density and velocity in the wind region are anti-correlated (bottom and middle panels of Figure \ref{1Msun_structures}); for example, $L_\mathrm{K,out}$ of M1 is slightly larger than $L_\mathrm{K,out}$ of M0 in spite of the smaller $L_\mathrm{A,tc}$ and $\dot{M}$ as the larger $v_r$ compensates the smaller $\rho$ in $L_\mathrm{K,out}$. 

The density at the coronal base also controls the energy loss from the corona. We are presenting radiative and gravitational losses evaluated at $r=r_\mathrm{tc}$ in Table \ref{energy balances}, where the integration for $L_\mathrm{R}$ is taken from $r=r_\mathrm{tc}$ to $r_\mathrm{out}$. We note that $L_{\rm G,tc}$ is exactly proportional to the density at $r=r_\mathrm{tc}$ (see equation \ref{LG}) and that $L_\mathrm{R}$ practically includes the conductive loss, $L_\mathrm{c}$, because the downward conductive flux from the corona to the chromosphere radiates away \citep{rosner1978, washinoue2023}. Since the radiative cooling is proportional to $\rho^2$ in the optically thin corona (equation \ref{thin cooling}), higher coronal density enhances $L_\mathrm{R,tc}$. Therefore, M0 gives the largest $L_\mathrm{R,tc}$ among the four cases, M0--M3.

The comparison between M0 and M3 indicates that the non-ideal MHD effects reduce the mass loss rate $\dot{M}$ by a factor of 6. $\dot{M}$ of the ideal MHD case, M0, is calibrated to explain the observational value $\approx$ $2\times 10^{-14}~\rm M_\odot ~yr^{-1}$ \citep{withbroe1988, wood2005, wood2021}, indicating that the cases with ambipolar diffusion (M2 and M3) cannot reproduce the average $\dot{M}$ of the current solar wind (Table \ref{energy balances}). However, we would like to note that there are still a number of freedoms in our setup; we particularly focus on the effect of the velocity perturbation at the photosphere on the global properties of the wind in Section \ref{subsec:delta v0}.


\subsection{Dissipation and Reflection of Transverse Waves\label{subsec:waves}}

\begin{figure}
 \centering
 \includegraphics[scale=0.7]{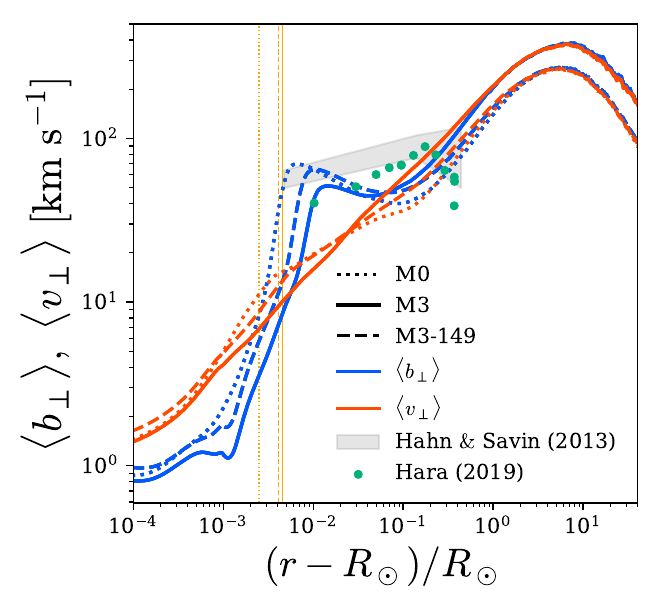}
 \caption{Time-averaged rms magnetic (blue) and velocity (red) amplitudes of transverse fluctuations for M0 (dotted), M3 (solid), and M3-149 (dashed). The altitude where $T=2\times 10^4$ K is plotted by orange vertical lines. The shaded region and green circles are observed nonthermal broadening by \citet{hahn2013} and \citet{hara2019}, respectively.}
 \label{Bpvp 6lines}
\end{figure}

Figure \ref{Bpvp 6lines} compares the time-averaged and root-mean-squared (rms) amplitudes of magnetic, $\langle b_\perp\rangle$ ($\equiv\sqrt{\langle B_\perp^2\rangle}/\sqrt{4\pi \langle\rho\rangle})$, (blue) and velocity, $\langle v_\perp\rangle (=\sqrt{\langle v_\perp^2 \rangle})$, (red) amplitudes for M0 (dotted) , M3 (solid), and M3-149 (dashed; see Section \ref{subsec:sameMdot}). 
These cases show $\langle v_\perp\rangle > \langle b_\perp\rangle$ in the chromosphere; particularly in the cases with magnetic diffusion $\langle b_\perp\rangle$ is decreased at $r-R_\odot \approx 10^{-3}R_\odot$ in the upper chromosphere owing to ambipolar diffusion (equation \ref{induction}). However, $\langle b_\perp\rangle$ rapidly increases in the transition region, leading to $\langle b_\perp\rangle > \langle v_\perp\rangle$ in the low corona. This indicates that the magnetic fluctuation behaves in a sense to conserve $B_\perp (=b_\perp\sqrt{4\pi\rho})$ across the transition region with a huge density gap \citep[][see also \citet{grappin2008}]{verdini2012}.

\begin{figure}
 \centering
 \includegraphics[scale=0.7]{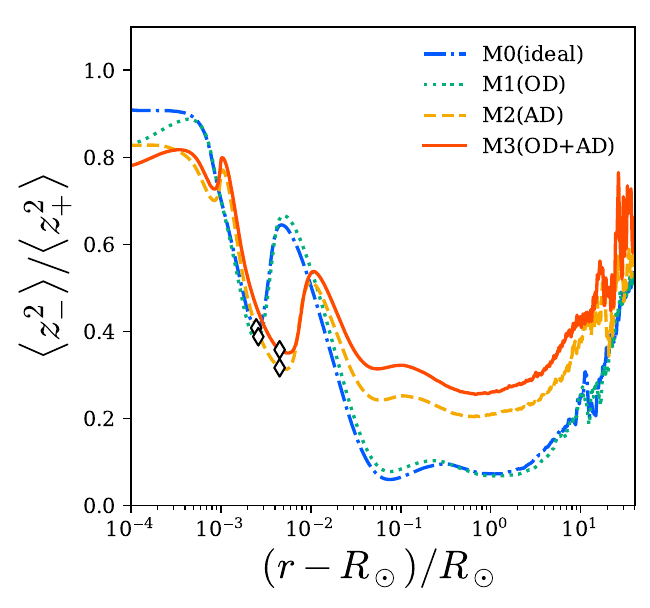}
 \caption{Time-averaged radial profile of the Alfv\'en ratio $(=\langle z_{-}^2\rangle /\langle z_{+}^2\rangle)$, where the line styles are the same as in Figure \ref{xeandRm}. 
 Diamond markers represent the location of the top of the chromosphere at $T=2\times 10^{4}~\text{K}$.}
 \label{Els,ref}
\end{figure}

The inequality between $\langle v_\perp\rangle$ and $\langle b_\perp\rangle$ in the low atmosphere reflects the fact that the transverse perturbations are not in a simple Alfv\'enic state but the injected outgoing Alfv\'en waves are substantially reflected. Wave reflection occurs mainly because of the variation in the Alfv\'en speed \citep{hollweg1984, an1990, suzuki2006, shoda2016} and of the field line curvature \citep{li2007}.
The wave reflection is the primary reason why the only tiny fraction of the input energy, $L_{\text{A},+,0}$, can contribute to the kinetic energy of the solar wind (Table \ref{energy balances} and Section \ref{subsec:structures}). To inspect the detailed properties of the reflection, 
Figure \ref{Els,ref} compares Els\"asser ratio, $\cal{R}_\mathrm{E}$ $\equiv \langle z_{-}^2\rangle / \langle z_{+}^2\rangle$ of M0 (dash-dotted), M1 (dotted), M2 (dashed), and M3 (solid). 
From the photosphere to the low chromospheric region, $r-R_\odot \lesssim 10^{-3}R_\odot$, $\cal{R}_\mathrm{E}$ is smaller in diffusive cases. This is because reflected waves, which have traveled a longer distance at a given $r$ than the outgoing waves coming directly from the photosphere, are more severely damped by non-ideal MHD effects.
The location of the local peak in $\cal{R}_\mathrm{E}$ around $10^{-2} R_\odot$ coincides with the transition region where the Alfv\'en velocity most drastically changes owing to the drop in the density. The peak value of $\cal{R}_\mathrm{E}$ is smaller in M2 and M3 with ambipolar diffusion because the density drop at the transition is smaller at the transition region (middle panel of Figure \ref{1Msun_structures}), which is due to the smaller temperature jump (top panel) as a result of the suppressed chromospheric evaporation (Section \ref{subsec:structures}). 

The lower coronal temperature due to the suppressed chromospheric evaporation also leads to the faster decrease of the coronal density as the pressure scale height is smaller. As a result, more efficient reflection takes place in the corona and wind regions of M2 and M3 to give larger $\cal{R}_\mathrm{E}$. In other words, the non-ideal MHD effects in the chromosphere indirectly reduce the energy transport by Alfv\'enic waves in the corona through the promoted wave reflection. However, even in M2 and M3, $\cal{R}_\mathrm{E}$ is still not large $\lesssim 0.4$ in $r-R_\odot\lesssim 10 R_\odot$, namely the Alfv\'enic Poynting flux is dominated by the outgoing component, being in $\langle v_\perp\rangle \approx \langle b_\perp\rangle$ as shown in Figure \ref{Bpvp 6lines}.


\subsection{Dependence on $\langle \delta v_0 \rangle$}\label{subsec:delta v0}
So far we have fixed the velocity perturbation at the photosphere to $\langle \delta v_0 \rangle = 1.25$ km s$^{-1}$. While this is a typical value as discussed in Section \ref{subsec:boundary condition}, observational data exhibit a reasonably large range. For example, \citet{oba2020} reviewed that horizontal convective velocities by various observations are ranging from $0.37~\rm km~s^{-1}$ to $2.4~\rm km~s^{-1}$. In this subsection, we investigate the dependence of the structure of the atmosphere and wind on $\langle \delta v_0 \rangle$. We magnify both transverse and longitudinal perturbations simultaneously, and perform simulations of M3 considering both Ohmic and ambipolar diffusion.

\begin{figure}
 \centering
 \includegraphics[scale=0.75]{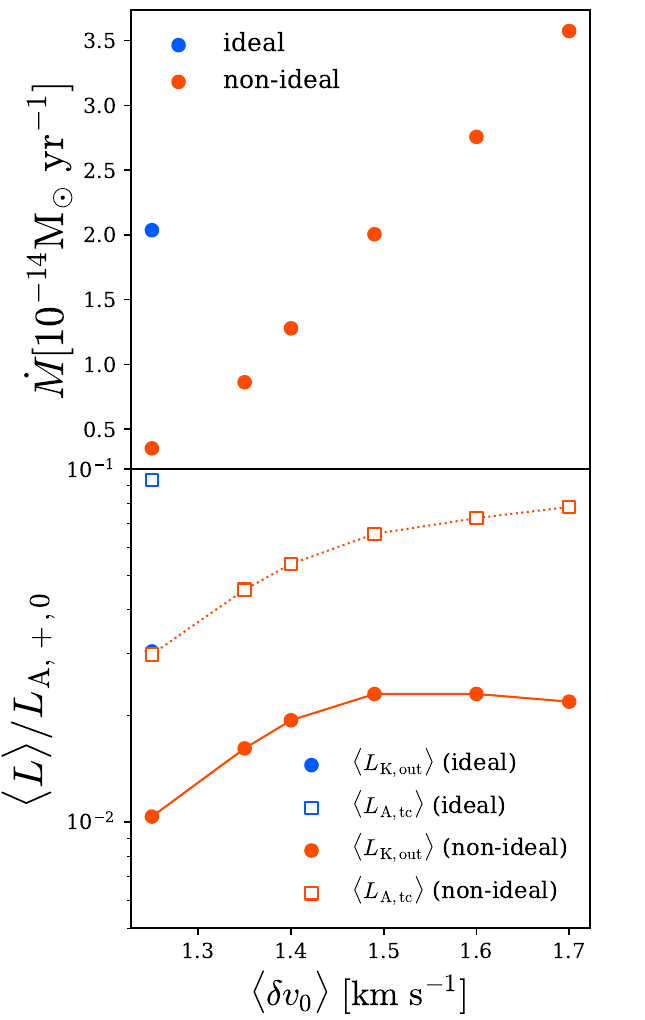}
 \caption{
  Dependence of $\dot{M}$ (filled circles in top panel), $L_\text{A,tc}/L_{\text{A,+,0}}$ (open squares in bottom panel), and  $L_\text{K,out}/L_{\text{A,+,0}}$ (filled circles in bottom panel) on $\langle \delta v_0 \rangle$. The red and blue symbols denote the results with both Ohmic and ambipolar diffusion (M3-*) and without magnetic diffusion (M0), respectively.
 }
 \label{1Msun_amps}
\end{figure}

\begin{figure}
 \centering
 \includegraphics[scale=0.65]{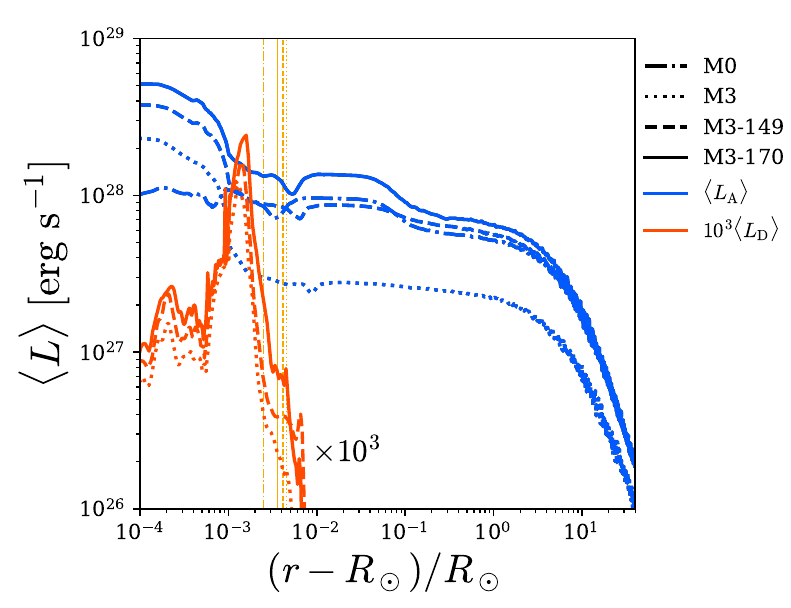}
 \caption{Time-averaged $L_\mathrm{A}$ (blue) and $L_\mathrm{D}$ (red) for M0 (dash-dotted), M3 (dotted), M3-149 (dashed), and M3-170 (solid). $L_\mathrm{D}$ is multiplied by $10^3$, and the vertical orange lines denote the locations at $T=2\times 10^4$K.}
 \label{dv LA LD}
\end{figure}

Figure \ref{1Msun_amps} presents the mass loss rate (top) and energy efficiencies (bottom) against $\langle \delta v_0 \rangle$. One may find that $\dot{M}$ drastically increases with $\langle \delta v_0 \rangle$ (red filled circles); by changing $\langle \delta v_0 \rangle$ from  1.25 km s$^{-1}$ to 1.49 km s$^{-1}$, $\dot{M}$ is enhanced by six times to reproduce the level of the current solar wind 
\citep{withbroe1988}. This sensitive dependence arises from the increasing trend of the survival fraction of the Alfv\'enic Poynting flux, $L_\mathrm{A,tc}/L_\mathrm{A,0}$, at the transition region (open squares in the bottom panel). 

\begin{figure}
  \begin{minipage}{0.45\linewidth}
    \includegraphics[scale=0.7]{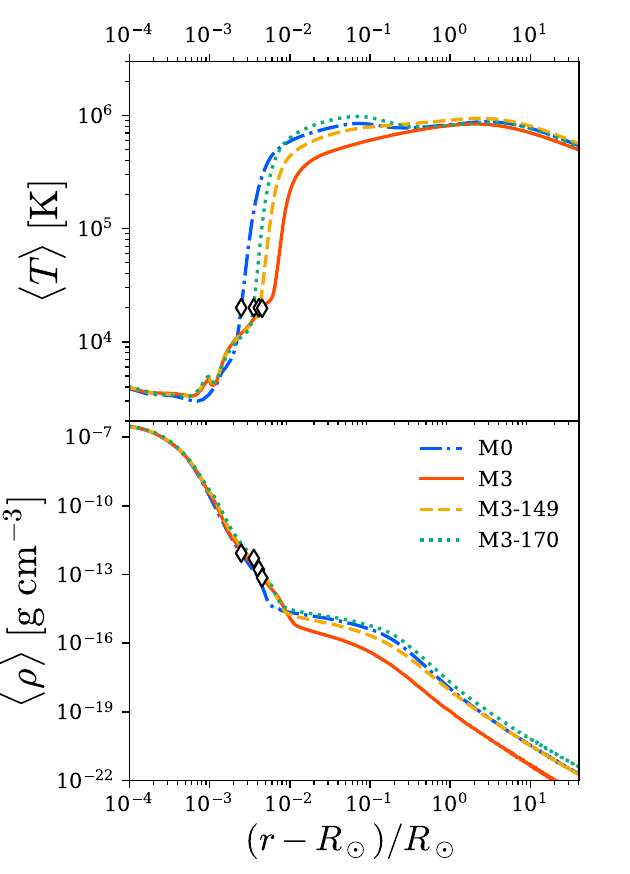}
  \end{minipage}\\
  \begin{minipage}{0.45\linewidth}
    \includegraphics[scale=0.7]{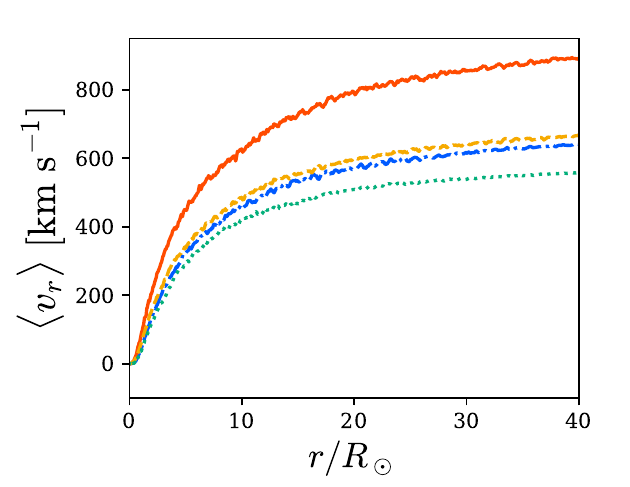}
  \end{minipage}
  \caption{The same as Figure \ref{1Msun_structures} but for M0 (blue dash-dotted), M3 (red solid), M3-149 (orange dashed), and M3-170 (green dotted). In the bottom panel for $\langle v_r \rangle$, the linear scale, $r/R_\odot$, is adopted for the horizontal axis.}
  \label{dv_1Msunstructures}
\end{figure}

To examine the radial variation of the Poynting flux, we compare $L_\text{A}$ (blue) and $L_\text{D}$ (red) of three non-ideal MHD cases (M3) with different $\langle \delta v_0 \rangle$ and the ideal MHD case (M0) in
Figure \ref{dv LA LD}. The qualitative trend of the efficient ambipolar dissipation in the chromosphere is similar in these three cases. However, a close look reveals that, although the dissipative Poynting luminocity, $L_\mathrm{D}$, is larger for cases with larger $\langle \delta v_{\perp,0}\rangle$, the difference among the three cases is not as large as that of  $L_\mathrm{A}$. This is because the ambipolar diffusion is less efficient in denser gas (equation \ref{eAD}). The middle panel of Figure \ref{dv_1Msunstructures} shows that the density in the chormosphere is highest in the case with the largest $\langle \delta v_{\perp,0}\rangle$, M3-170 (green dotted line), as the gas is supported by the magnetic pressure, $B_\perp^2/8\pi$, associated with Alfv\'enic perturbations (blue dashed line in Figure \ref{Bpvp 6lines}), in addition to the gas pressure. As a result, the ambipolar diffusion is relatively quenched in this case, compared to that expected from the simple extrapolation from cases with smaller $\langle \delta v_{\perp,0}\rangle$. Therefore, the original case, M3, with the smallest $\langle \delta v_{\perp,0}\rangle$ suffers the severest ambipolar damping in dimensionless units, $L_\mathrm{D}/L_\mathrm{A}$, in the chromosphere.  
Additionally, the slower decrease of the density in the choromosphere suppresses the reflection of Alfv\'enic waves in cases with large $\langle \delta v_{\perp,0}\rangle$ \citep[][see also, Section \ref{subsec:waves}]{suzuki2006,suzuki2013}.
These are the reasons why the survival fraction, $L_\mathrm{A,tc}/L_\mathrm{A,+,0}$, at the transition region increases with $\langle \delta v_{\perp,0}\rangle$ in the bottom panel of Figure \ref{1Msun_amps}.

The kinetic energy luminocity, $L_\text{K,out}/L_{\text{A,}+,0}$, exhibits a similar trend to $L_\text{A,tc}/L_{\text{A,}+,0}$, but it is slightly decreasing with $\langle \delta v_0 \rangle$ for $\langle \delta v_0 \rangle >1.49~\rm km~s^{-1}$. This stems from enhanced radiative cooling (Table \ref{energy balances}), which is augmented by the increased coronal density (middle panel of Figure \ref{dv_1Msunstructures}); the larger $L_\mathrm{A,tc}$ heats up the corona to higher temperature (top panel), which promotes chromospheric evaporation. The higher coronal density yields smaller wind velocity (bottom panel of Figure \ref{dv_1Msunstructures}), which is also a reason for the saturated $L_\text{K,out}/L_{\text{A,}+,0}$. We note that both $L_\text{A,tc}/L_{\text{A,}+,0}$ and $L_\text{K,out}/L_{\text{A,}+,0}$ are smaller than those of the ideal MHD case with $\langle \delta v_0 \rangle = 1.25$ km s$^{-1}$, M0 (blue points in Figure \ref{1Msun_amps}) within the range of $\langle \delta v_0 \rangle \le 1.70$ km s$^{-1}$.

In Figure \ref{dv_1Msunstructures} we are also displaying the result of the ideal MHD case (M0; blue dashed lines) to compare with that of the cases with magnetic diffusion. One can see that the diffusive case with $\langle \delta v_0 \rangle=1.49$ km s$^{-1}$ (M3-149; orange dashed lines) almost traces the ideal case with $\langle \delta v_0 \rangle=1.25$ km s$^{-1}$ in $r-R_{\odot}\gtrsim 0.5R_\odot$ and gives the comparable $\dot{M}$ (Table \ref{energy balances}); the difference between these two cases is seen only below the low corona.

The same tendency is obtained for the amplitudes of Alfv\'enic waves (Figure \ref{Bpvp 6lines}). Both magnetic and velocity amplitudes of M3-149 (dashed lines) almost coincide with those of M0 (dotted lines) above $r-R_\odot\gtrsim R_\odot$. Paradoxically, $\langle b_\perp\rangle$ and $\langle v_\perp\rangle$ of M3-149 are smaller than those of M3 with smaller $\langle \delta v_0 \rangle=1.25$ km s$^{-1}$ there (solid lines) even though the Alfv\'enic Poynting luminocity is larger (Figure \ref{dv LA LD}). This is because the density is higher by nearly an order of magnitude (middle panel of Figure \ref{dv_1Msunstructures}). In other words, the higher coronal density by the boosted chromospheric evaporation can transport larger $L_\mathrm{A}$($\propto \rho v_\perp^2\sim \rho b_\perp^2$) to the outer region with smaller magnetic and velocity amplitudes. 


\section{Discussions} \label{sec:discussion}

\subsection{Density Fluctuation} \label{subsec:sameMdot}
\begin{figure}
 \centering
 \includegraphics[scale=0.7]{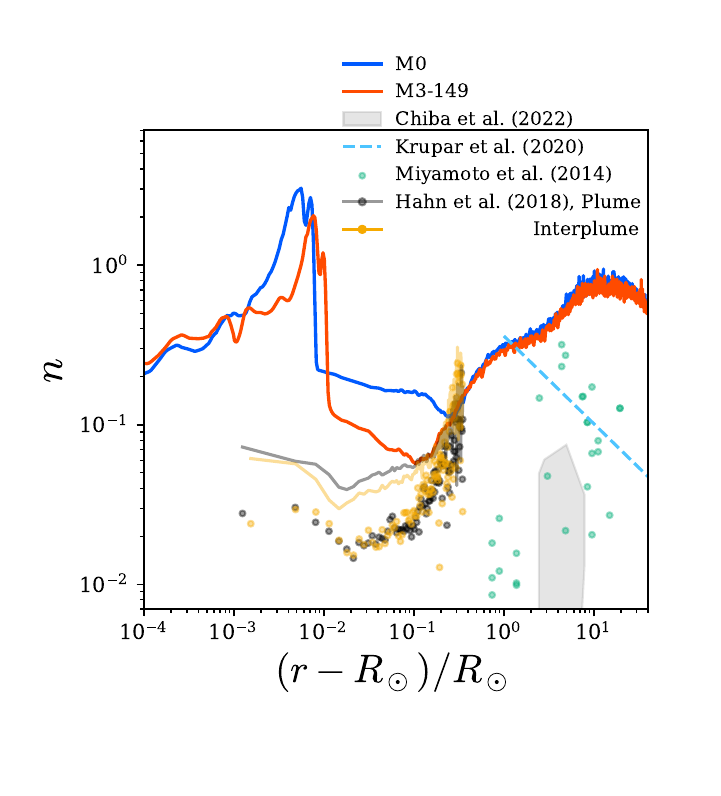}
 \caption{Time-averaged radial profile of relative density fluctuation, $n$, (equation \ref{dfluc}) of M0 (blue solid line) and M3-149 (red solid line). Light blue-dashed line represents the power-law fit to observed $n$ to explain type III radio bursts by the PSP \citep{krupar2020}, green dots and gray area are derived from radio scintillations by Akatsuki \citep{imamura2014, miyamoto2014, chiba2022}, and orange and black dots and lines are obtained from fluctuations of coronal intensity detected by PROBA2/SWAP \citep[][see text for the detail]{hahn2018}.}
 \label{rho fluctuation}
\end{figure}

In Section \ref{subsec:delta v0}, we demonstrated that the non-ideal MHD case $\langle \delta v_0 \rangle = 1.49~\rm km~s^{-1}$, M3-149, and the ideal MHD case with $\langle \delta v_0 \rangle = 1.25~\rm km~s^{-1}$, M0, give similar corona and wind properties with $\dot{M}$ being comparable to the mass loss rate of the present-day solar wind. However, as the propagation and dissipation of Alfv\'enic waves below the transition region are different between these two cases (Figure \ref{dv LA LD}), we expect that there would be observational footprints to grab the effects of the magnetic diffusion in the low atmospheric region.
As a potential candidate for such observational signatures, we examine the radial profiles of dimensionless density fluctuation,
\begin{equation}
    \label{dfluc}
    n = \frac{1}{\langle \rho \rangle}\sqrt{\langle (\rho - \langle \rho \rangle)^2 \rangle},
\end{equation}
in Figure \ref{rho fluctuation}.

In the chromosphere, $r-R_{\odot}<10^{-2}R_{\odot}$, both cases exhibit large density fluctuations, which are mainly due to the longitudinal waves generated at the photosphere. Their amplitudes are amplified as propagating upward in the density-decreasing atmosphere. However, these longitudinal waves are steepen to form shocks and dissipated before reaching the corona (Section \ref{subsec:structures}). Therefore, the density fluctuations above the transition region are excited in situ.

One may find a remarkable difference in $n$ in the low coronal region, $10^{-2}R_\odot\lesssim r-R_\odot\lesssim 0.3R_\odot$. These density fluctuations are excited by the variation in magnetic pressure, $B_\perp^2/8\pi$, with transverse waves \citep{hollweg1982, kudoh1999, suzuki2005} and parametric decay instability \citep{goldstein1978, terasawa1986, suzuki2006}. In the non-ideal MHD case, higher frequency Alfv\'enic waves are preferentially damped by the ambipolar diffusion in the chromosphere \citep{DePontieu1998, leake2005, zaqarashvili2011}. As a result, only lower-frequency waves survive beyond the upper chromosphere. Hence, small-scale density fluctuations are not excited, which reduces $n$ in the low corona. 
However, as small-scale structures are regenerated in the corona by the interaction between pre-existing outgoing waves and reflected waves from the upper region, $n$'s of the two cases are converged in $r-R_\odot\gtrsim 0.3 R_\odot$. The peak of $n$ at $r-R_\odot \approx 10R_\odot$ is formed by the decay instability \citep{shoda2018b}. 

We also plot observed density fluctuations by various instruments in Figure \ref{rho fluctuation}. The numerical results exceed most of the observational data. On the one hand, MHD simulations in a 1D flux tube generally tend to overestimate density perturbations because compressible waves are confined in the tube. On the other hand, the density fluctuations estimated from radio scintillation measurements with the Akatsuki spacecraft by \citet[][green dots; see also \citet{imamura2014}]{miyamoto2014}; \citet[][gray shade]{chiba2022} might be underestimated if positive and negative density fluctuations were partially cancelled out along the line of sight. The similar effect may also affect the density fluctuations obtained from type III radio bursts using the Radio Frequency Spectrometer onboard the Parker Solar Probe \citep[][light blue dashed line; see also \citet{Kontar2023,Krupar2024}]{krupar2020}. Observed density fluctuations in the low corona can be derived from coronal intensity variation by the the Sun Watcher using the Active Pixel System detector and Image Processing on the Project for Onboard Autonomy (SWAP/PROBA2) \citep[][orange and black dots and lines]{hahn2018}, where black and orange dots (lines) are obtained with a running- (average-)difference method in plume and interplume regions, respectively; the running difference approach gives more or less an "average" level of the fluctuations, while the average difference one gives an upper bound that may include spectral changes in low frequency parts. The comparison with the simulation results shows that these observational data favor the case with the non-ideal MHD effects (M3-149; red line).

\subsection{3-dimensional Magnetic Diffusion}\label{hall effects}
We have ignored the Hall term and the part of ambipolar diffusion that requires the nonlinear coupling of both transverse components of magnetic field (Appendix \ref{appendix}). Since these terms stem from drifts of particles and excite magnetic fluctuations from one component to the other, they are three dimensional processes. 
When the direction of wave propagation is not parallel with the magnetic field, Hall and ambipolar drifts may cause unstable phenomena; while ambipolar drifts may destabilize obliquely propagating waves, Hall instability possibly amplifies both perpendicular and oblique waves \citep{desch2004, pandey2012, pandey2013}.

A typical example of the Hall instability occurs when radial shear flow generates an azimuthal magnetic field from a radial field. If the Hall drift is active, the radial magnetic field is amplified from the generated azimuthal field; the magnetic fields of the different components are amplified each other \citep{pandey2008}. This instability possibly occurs in the chromosphere when torsional \Alfven waves are excited by vortex motion in the photosphere \citep{fedun2011,iijima2017,srivastava2017,kuniyoshi2023}. 

Since Hall instability makes open flux tubes unstable, it may have a huge impact on our model based on the flux open flux tube, equation (\ref{expansion factor}). Although the Hall term can be included in the 1D system of this study (Appendix \ref{appendix}), this treatment is insufficient because the only small attacking angle between wave and magnetic field is allowed; the direction of wave propagation is strictly fixed along $r$ and the direction of magnetic field is deviated from $r$ only by $B_\perp/B_r$. The influence of Hall instability should be investigated by a multidimensional numerical model.

\subsection{Non-equilibrium Ionization}
\label{subsec:noneq_ion}
We are assuming the ionization and recombination are in equilibrium when calculating the ionization degree, which is a critical parameter in determining the magnetic diffusivities (Section \ref{subsec:boundary condition}). While this assumption is reasonable in the photosphere, non-equilibrium ionization plays an important role in the chromosphere where the timescale to achieve ionization equilibrium could be longer than a typical MHD timescale \citep{leenaarts2020}. For example, the equilibration timescale for hydrogen to balance ionization and recombination is as long as $10^5$s in the mid- to upper chromosphere \citep{carlsson2002}, which is longer than the transit timescale $\sim 10^2 - 10^3$ s of Alfv\'en waves traveling across the chromosphere. The equilibration timescale for helium in the upper chromosphere and the transition region is $\sim 10^2-10^3$ s \citep{Golding2014}, which is shorter but still comparable to the periods of the Alfv\'en transit timescale.

These arguments indicate the importance of the non-equilibrium effects in the magnetic diffusion. It is inevitable to take into account in more elaborated studies, although it is computationally expensive to properly handle them in numerical simulations \citep[e.g.,][]{leenaarts2006,leenaarts2020}.

\subsection{Low-mass Main-sequence Stars}
The Alfv\'en-wave driven mechanism is believed to be also a promising process in driving stellar winds from low-mass main sequence stars \citep{cranmer2011,sakaue2021a,sakaue2021b,wood2021}. Compared to the Sun, the non-ideal MHD effects are probably more essential in these stars because the temperature in the photosphere is lower. 

\section{Summary\label{sec:summary}}

We investigated the influence of non-ideal MHD effects on the MHD-wave-driven solar wind by performing 1D non-ideal MHD simulations with radiative cooling and thermal conduction.
In the photosphere and the choromosphere the plasma is partially ionized (top panel of Figure \ref{xeandRm}) so that the non-ideal MHD effects play a significant role. The radial profile of magnetic Reynolds number (bottom panel of Figure \ref{xeandRm}) indicates that Ohmic diffusion is non-negligible from the photosphere to the low chomosphere and that ambipolar diffusion is substantially important in the chromosphere.

The magnetic-field fluctuations of Alfv\'enic waves from the photosphere are significantly damped by ambipolar diffusion in the chromosphere (Figures \ref{LAandLD} and \ref{Bpvp 6lines}), reducing the Poynting flux that reaches the corona (Figure \ref{LA pms}). As a result, the coronal temperature is lower than that obtained in the ideal MHD simulation, which suppresses the chromospheric evaporation and reduces the coronal density (Figure \ref{1Msun_structures}). Consequently, the mass loss rate of the model with Ohmic and ambipolar diffusion is reduced by a factor of 6, compared with that of the ideal case (Table \ref{energy balances}).
The coronal density also decreases more rapidly with height owing to the lower coronal temperature, and hence, a larger fraction of the outgoing Alfv\'enic waves is reflected to give higher Els\"asser ratio in the corona because of the larger gradient of the Alfv\'en velocity (Figure \ref{Els,ref}).

We also found that the physical properties of the corona and wind sensitively depends on the amplitude of velocity fluctuations, $\langle \delta v_0 \rangle$, at the photosphere. When $\langle \delta v_0 \rangle$ is increased from our standard value, 1.25 km s$^{-1}$ to 1.49 km s$^{-1}$, which corresponds to the increase of the input energy ($\propto \langle \delta v_0^2\rangle$) by $\approx 40$\%, the mass loss rate is enhanced to six times the original value to recover the mass loss rate obtained in the present-day solar wind (top panel of Figure \ref{1Msun_amps}). 
This is firstly because ambipolar dissipation is quenched in the higher-density chromosphere (Figure \ref{dv LA LD}) and secondly because the reflection of Alfv\'enic waves is suppressed. As a consequence, a larger fraction of the Alfv\'enic Poynting flux injected from the photosphere is transported to the corona (bottom of Figure \ref{1Msun_amps} and Table \ref{energy balances}), resulting in hotter corona and denser wind (Figure \ref{dv_1Msunstructures}).

The non-ideal MHD case with $\langle \delta v_0 \rangle=1.49$ km s$^{-1}$ and the ideal MHD case with $\langle \delta v_0 \rangle=1.25$ km s$^{-1}$ give similar structures of the corona and solar wind. 
However, the density fluctuation of the non-ideal MHD case is smaller in the low coronal region because ambipolar diffusion selectively damps high-frequency Alfv\'enic waves to quench the excitation of short-wavelength compressible perturbations by the parametric decay instability and the nonlinear mode conversion. Density perturbations in the corona can be used as an observational signature of the non-ideal MHD dissipation of MHD waves in the chromosphere.


\section*{Acknowledgments} 
We thank M. Hahn, T. Imamura, and V. Krupar for providing observational data and valuable comments.
Numerical computations were in part carried out on PC cluster at Center for Computational Astrophysics, National Astronomical Observatory of Japan. T.K.S. is supported by Grants-in-Aid for Scientific Research from the MEXT/JSPS of Japan, 22H01263. T.T. is supported by IGPEES, WINGS Program in the University of Tokyo and Research Fellowships for Young Scientists (JSPS KAKENHI Grant Number, 24KJ0605).

%

\vspace{5mm}

\appendix

\section{Hall and Ambipolar Diffusion Terms} \label{appendix}

The Hall and ambipolar diffusion terms in the induction equation (\ref{induction}) are written as

\begin{align}
\label{3Dvector induction}
    \pdv{\bm{B}}{t} &= \cdots + \nabla \cross \left[ -\eta_\mathrm{H}\qty(\nabla\cross \bm{B})\cross \bm{\hat{e}}_{B} \right.\\ \nonumber &+ \left. \eta_\text{AD}\qty(\qty(\nabla\cross \bm{B})\cross \bm{\hat{e}}_{B}) \cross \bm{\hat{e}}_{B}\right], 
\end{align}
where $\eta_\mathrm{H}\equiv \frac{c|\bm{B}|}{4\pi n_e e_\text{c}}$ is the Hall diffusivity and $\bm{\hat{e}}_{B}$ is the unit vector along a magnetic field line. In our 1D simulations with the coordinate system, equation (\ref{jouran vector}), the Hall part is explicitly expressed as
\\
\begin{equation}
      \left.\pdv{B_{\perp 1(2)}}{t}\right|_\text{H}= \pm \frac{1}{r\sqrt{f}}\pdv{r}\qty[\eta_\text{H}\pdv{r}\qty(B_{\perp 2(1)}r\sqrt{f})B_r],
      \label{eq:Hall_1D}
\end{equation}
which generates the first (second) transverse component from the second (first) component.  
We ignore this term although in a strictly speaking even our "$1\frac{2}{2}$" coordinate system can consider it.

For the ambipolar diffusion part, we have 
\begin{align}
    &\left.\pdv{B_{\perp 1(2)}}{t}\right|_\text{AD}= \frac{1}{r\sqrt{f}}\pdv{r} \left[\eta_\text{AD}\pdv{r} \left(B_{\perp 1(2)}r\sqrt{f}\right)\right. \nonumber \\
    &+ \left.\frac{\eta_\text{AD}}{B^2} \left\{\pdv{r}( B_{\perp 2(1)}r\sqrt{f})B_{\perp 1}B_{\perp 2} -\pdv{r}(B_{\perp 1(2)}r\sqrt{f})B_{\perp 2(1)}^2 \right\}\right]. 
    \label{eq:AD_1D2}
\end{align}
We only considered the first term on the right-hand side, which corresponds to isotropic diffusion. We ignore the other terms, which require nonlinear coupling between the two transverse components.

\bibliography{journal_cite}{}
\bibliographystyle{aasjournal}



\end{document}